\documentclass[acus]{JAC2003}

\usepackage{graphicx}
\usepackage{booktabs}

\newcommand{\be}{\begin{equation}}
\newcommand{\ee}{\end{equation}}
\newcommand{\bea}{\begin{eqnarray}}
\newcommand{\eea}{\end{eqnarray}}
\newcommand{\bi}{\begin{itemize}}
\newcommand{\ei}{\end{itemize}}
\newcommand{\ben}{\begin{enumerate}}
\newcommand{\een}{\end{enumerate}}

\newcommand{\lp}{\left(}
\newcommand{\rp}{\right)}

\def\frac#1#2{{{#1}\over {#2}}}
\def\gsim{\mathrel{\rlap{\lower4pt\hbox{\hskip1pt$\sim$}}
    \raise1pt\hbox{$>$}}}         
\def\lsim{\mathrel{\rlap{\lower4pt\hbox{\hskip1pt$\sim$}}
    \raise1pt\hbox{$<$}}}         

\newcommand{\draft}[1]{}

\def\beq{\begin{equation}}  
\def\eeq{\end{equation}}  

\def \n0{N_j^{(0)}}

\def\lapprox{\lower .7ex\hbox{$\;\stackrel{\textstyle <}{\sim}\;$}}
\def\gapprox{\lower .7ex\hbox{$\;\stackrel{\textstyle >}{\sim}\;$}}

\begin{document}
\title{Physics potential of precision
measurements of the LHC luminosity\thanks{This research 
has been supported by a Marie Curie Intra--European Fellowship
of the European Community's 7th Framework Programme under contract
number PIEF-GA-2010-272515.}}

\author{Juan Rojo\thanks{juan.rojo@cern.ch}, CERN PH/TH, Geneva, Switzerland}

\maketitle

\begin{abstract}
A precision measurement of the LHC luminosity is a
key ingredient for its physics program. 
In this contribution first of all we review the theoretical
accuracy in the computation of LHC benchmark processes.
Then we discuss the impact of available and future LHC
data in global analysis of parton distributions, with
emphasis on the treatment of luminosity uncertainties.
Finally we present some suggestions for the physics
opportunities that can become available
by measuring ratios of cross sections between
the 8 TeV and 7 TeV runs.

\end{abstract}

\section{THEORETICAL ACCURACY OF LHC STANDARD CANDLES}

\begin{figure}[htb]
   \centering
   \includegraphics*[width=75mm]{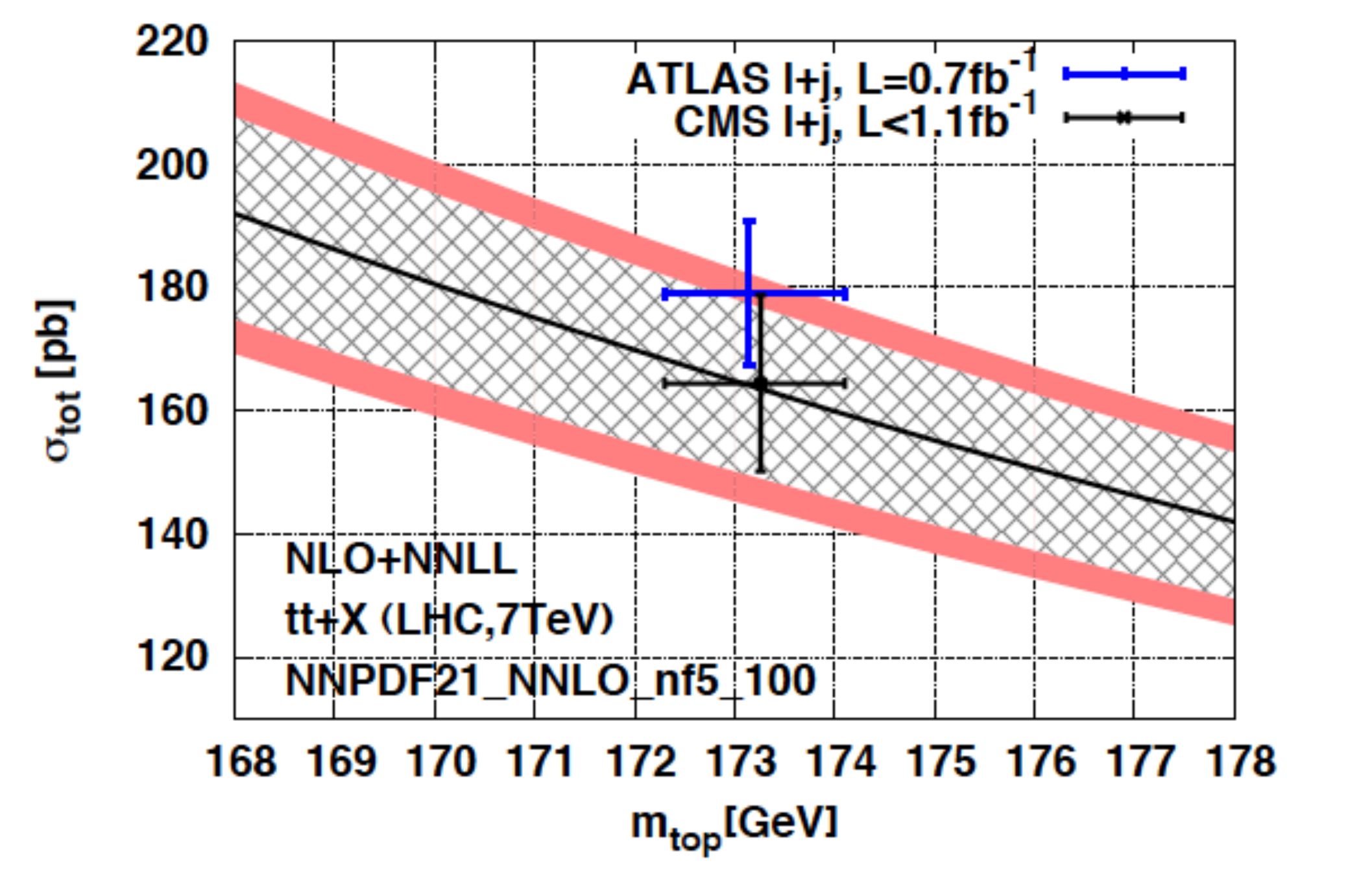}
   \caption{Comparison of the most updated theoretical
predictions for top quark pair production with ATLAS and CMS
data. Figure taken from~\cite{Cacciari:2011hy}. }
   \label{fig:top-mangano}
\end{figure}

An accurate determination of the luminosity uncertainty is
specially 
important when comparing data and theory for the total rates
of relevant processes like top quark or electroweak boson
production, for which the theoretical predictions are very
accurate.
For top quark pair production the theoretical accuracy is NLO+NNLL.
An up--to--date comparison between theoretical predictions and
experimental data for top quark pair cross sections
has been presented in Ref.~\cite{Cacciari:2011hy}.
The comparison of these updated
predictions with the most recent ATLAS and CMS data is show
in Fig.~\ref{fig:top-mangano}.
  The current accuracy in the theoretical computation
of the cross section is about 18 pb (10\%) from  higher orders and
PDF uncertainties. It is reasonable
to expect that this theory error will decrease down to 3-5 pb once
the full NNLO calculation and more updated PDFs with LHC data become
available. This would allow stringent tests of new
physics scenarios by comparing
to the LHC data, provided that luminosity uncertainties can be reduced
down to a similar level.

\begin{figure}[htb]
   \centering
   \includegraphics*[width=75mm]{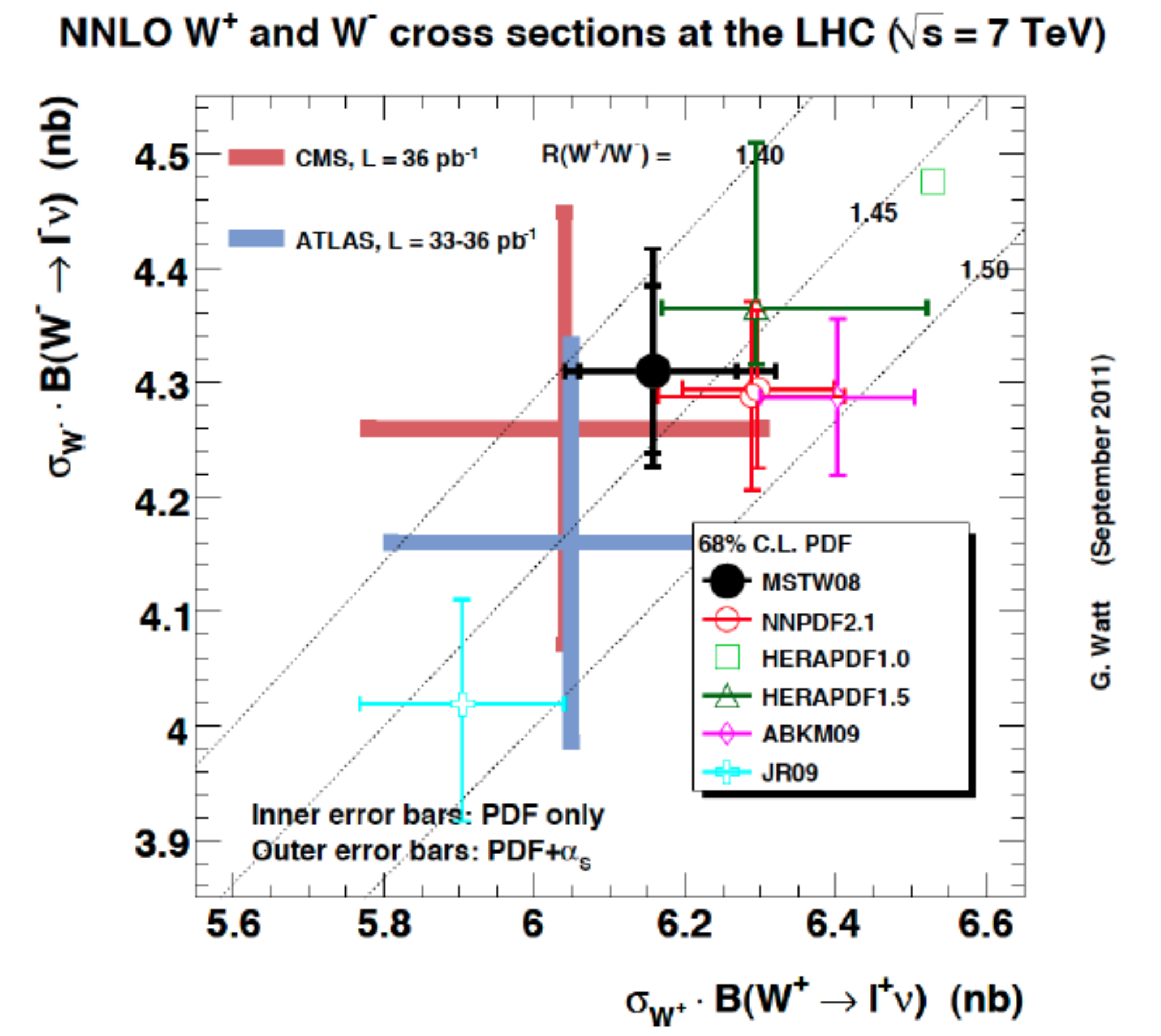}
   \caption{Comparison of the recent ATLAS and CMS
data on the $W^+$ and $W^-$ total cross sections
with predictions from different PDF sets. 
Figure taken from~\cite{Watt:2012fj}. }
   \label{fig:w-watt}
\end{figure}

Another very important cross sections are the $W$ and $Z$
cross sections. In particular, 
the $Z$ cross section has been used as a cross check
of the calibration of the absolute luminosity. 
The NNLO corrections for this processes are know since a
long time, and have recently been implemented in fully
differential programs like FEWZ~\cite{Gavin:2010az} or 
DYNNLO~\cite{Catani:2010en}
that allow to apply exactly the same
cuts as in the experimental analysis. In Fig.~\ref{fig:w-watt}
we show the comparison of the recent ATLAS and CMS
data on the $W^+$ and $W^-$ total cross sections
with predictions from different PDF sets. The comparison between
data and theory for these observables is now limited by
normalization uncertainties, so it is clear that more
stringent comparisons would be possible with an more accurate
determination of the LHC luminosity. This source of
uncertainty can be eliminated taking suitable ratios
of cross sections, like the $W^+/W^-$ ratio, but this
way useful information about the normalizations of the
distributions is lost.

\begin{figure}[htb]
   \centering
   \includegraphics*[width=80mm]{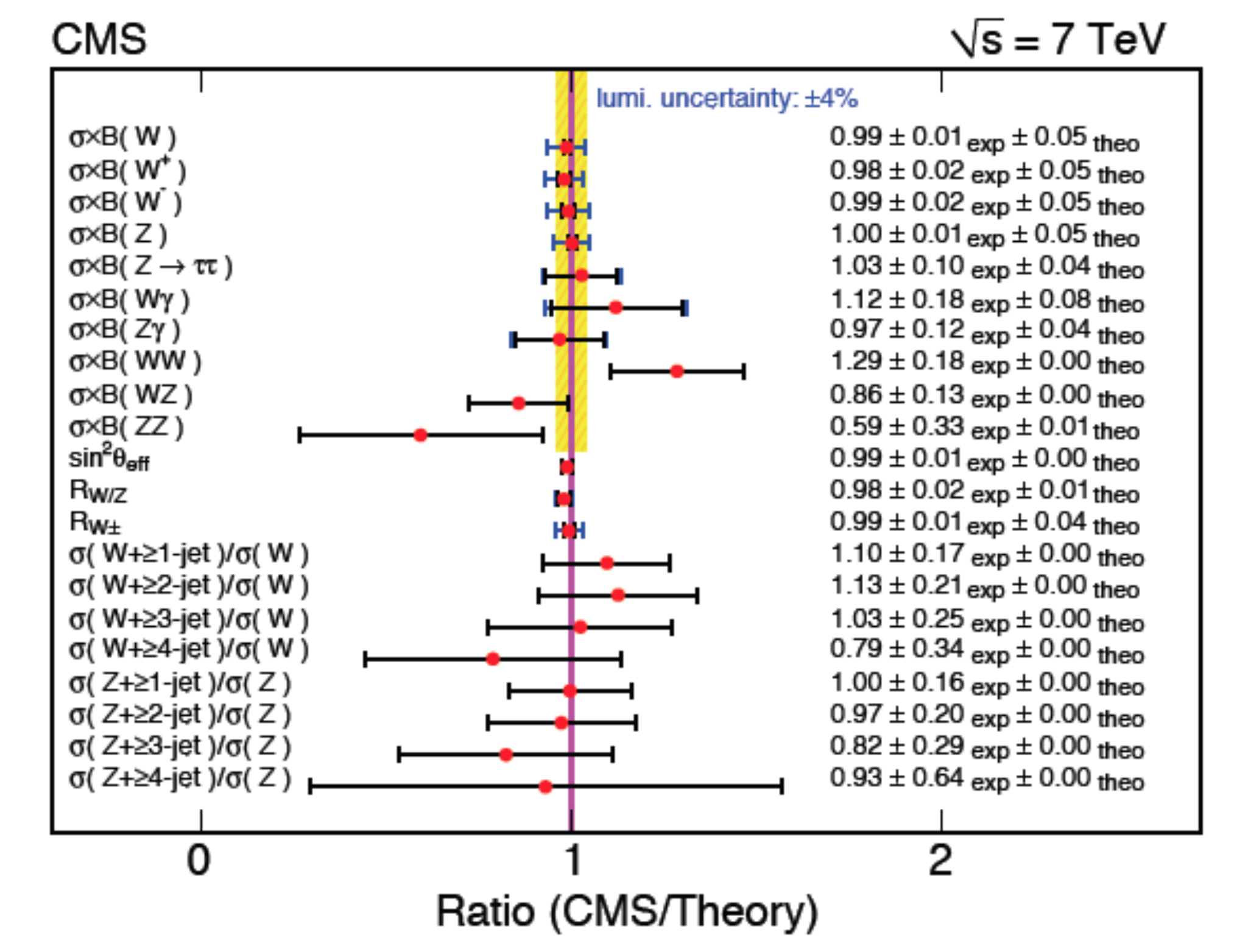}
   \caption{Comparison of a variety of
electroweak processes between theory and the
CMS measurements from the 2010 and 2011
data. The luminosity uncertainty of 4\% is also shown
}
   \label{fig:lep}
\end{figure}

$W,Z$ and top production are two of the best standard candles
at the LHC, but many other cross sections have been
measured with good accuracy to begin to challenge theory, and
is clear that at the level of an accuracy of 5-10\%, the
luminosity uncertainties begin to play a role. 
In Fig.~\ref{fig:lep} we show a recent compilation of CMS
electroweak measurements compared to theory predictions. 
One might expect that the accuracy of other processes like $WW$
or $Z\gamma $ to improve in the near future up to the point
in which they would benefit from a more accurate luminosity measurement.
Eventually, even processes like $W$ in association with one jet
could be measured precisely enough so that the comparison with theory
is limited by luminosity uncertainties.

All in all, there is a strong physics motivation to obtain
accurate determinations of the LHC luminosity to optimize
the comparisons between data and theory for several
standard candle processes. Now we turn to discuss the
impact of normalization errors and LHC data in the context of
global PDF analysis.

\section{IMPACT OF LHC DATA INTO PDFs}

The precise knowledge of parton distributions 
functions is an important requirement
for the LHC physics program. Just to mention two examples:
accurate PDFs, and in particular the gluon,
 are crucial to determine the theoretical
precision of the Higgs cross section in 
gluon fusion~\cite{Demartin:2010er,Dittmaier:2011ti},
and PDFs are now the dominant systematic error in the very
precise CDF determination of $M_W$~\cite{Aaltonen:2012bp}, 
which imposes tight constraints
on SM Higgs and new physics from indirect EW fits. In both cases
LHC data is known to improve the accuracy of the theoretical
predictions from the PDF side 
(see for example~\cite{Bozzi:2011ww} for the $M_W$ case.)

\begin{figure}[htb]
   \centering
   \includegraphics*[width=75mm]{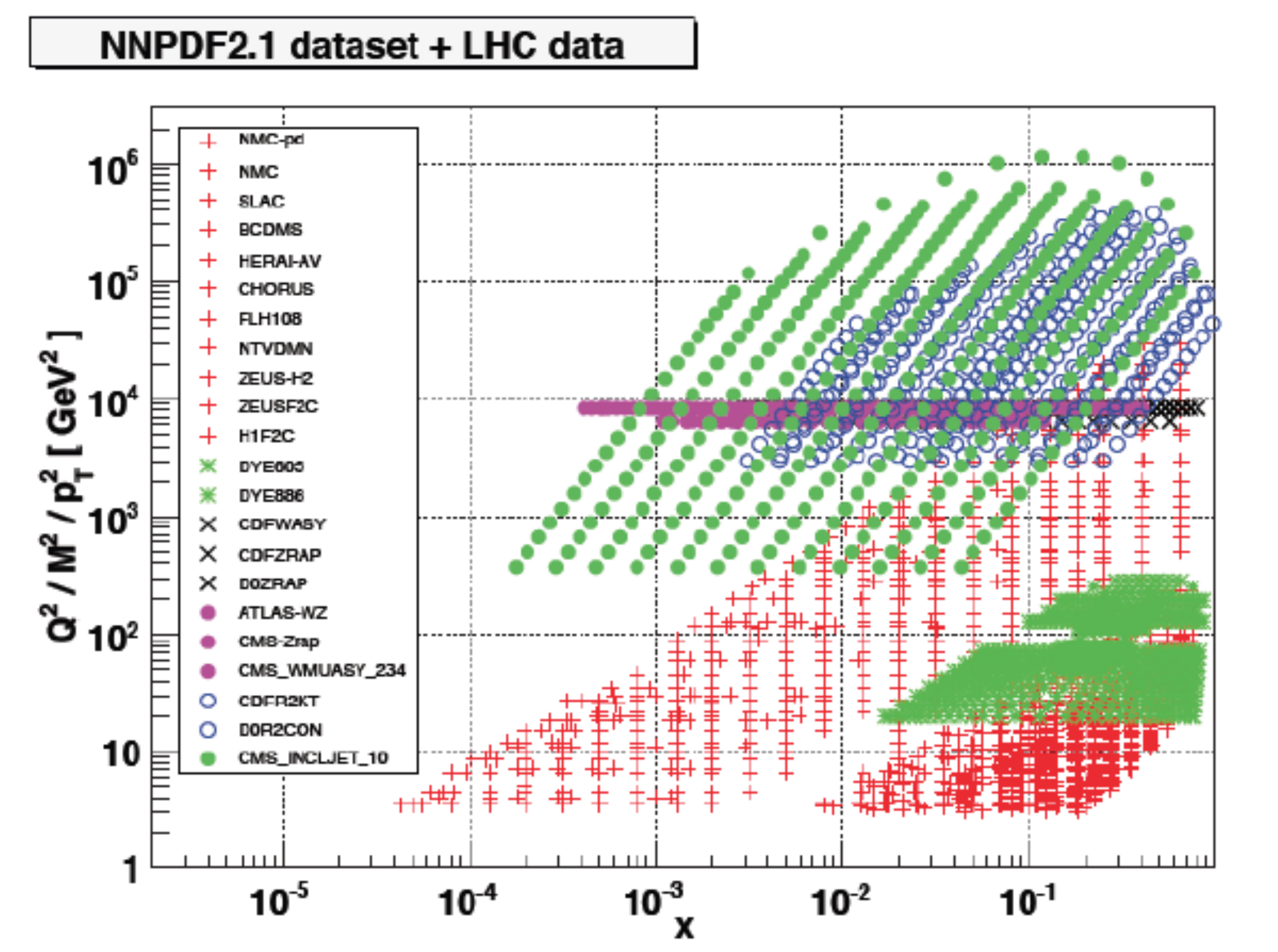}
   \caption{Kinematical coverage in the
$x,Q^2$ plane of the different experiments included
in the NNPDF2.1 analyses, supplemented with that of some
recent LHC measurements.
}
   \label{fig:kin}
\end{figure}

PDFs are determined from a wide variety of different data
like deep--inelastic scattering, Drell-Yan and jet production.
With the advent of the LHC, the emphasis is moving towards using
as much as possible solid and robust LHC data.
There are many processes that have been measured at the
LHC that can be used as an input to global PDF analyses, and many
others that will be measured in the next years. A necessarily incomplete list
is the following:
\begin{itemize}
\item Electroweak boson production, both inclusive and in association with jets and heavy quarks
\item Inclusive jet and dijet production
\item Isolated photon and photon+jet production
\item Top quark pair production distributions and single top production
\item Heavy flavor production
\end{itemize}
The kinematical coverage in the
$x,Q^2$ plane of the different experiments included
in the NNPDF2.1 analysis~\cite{Ball:2011mu}, supplemented with that of some
recent LHC measurements, is shown in Fig.~\ref{fig:kin}.
LHC measurements have already been used to constrain PDFs in
public sets:
for example, the NNPDF2.2 set~\cite{Ball:2011gg}  already 
includes the $W$ asymmetry data
obtained from 36 pb$^{-1}$ by ATLAS and CMS.

\begin{figure}[htb]
   \centering
   \includegraphics*[width=75mm]{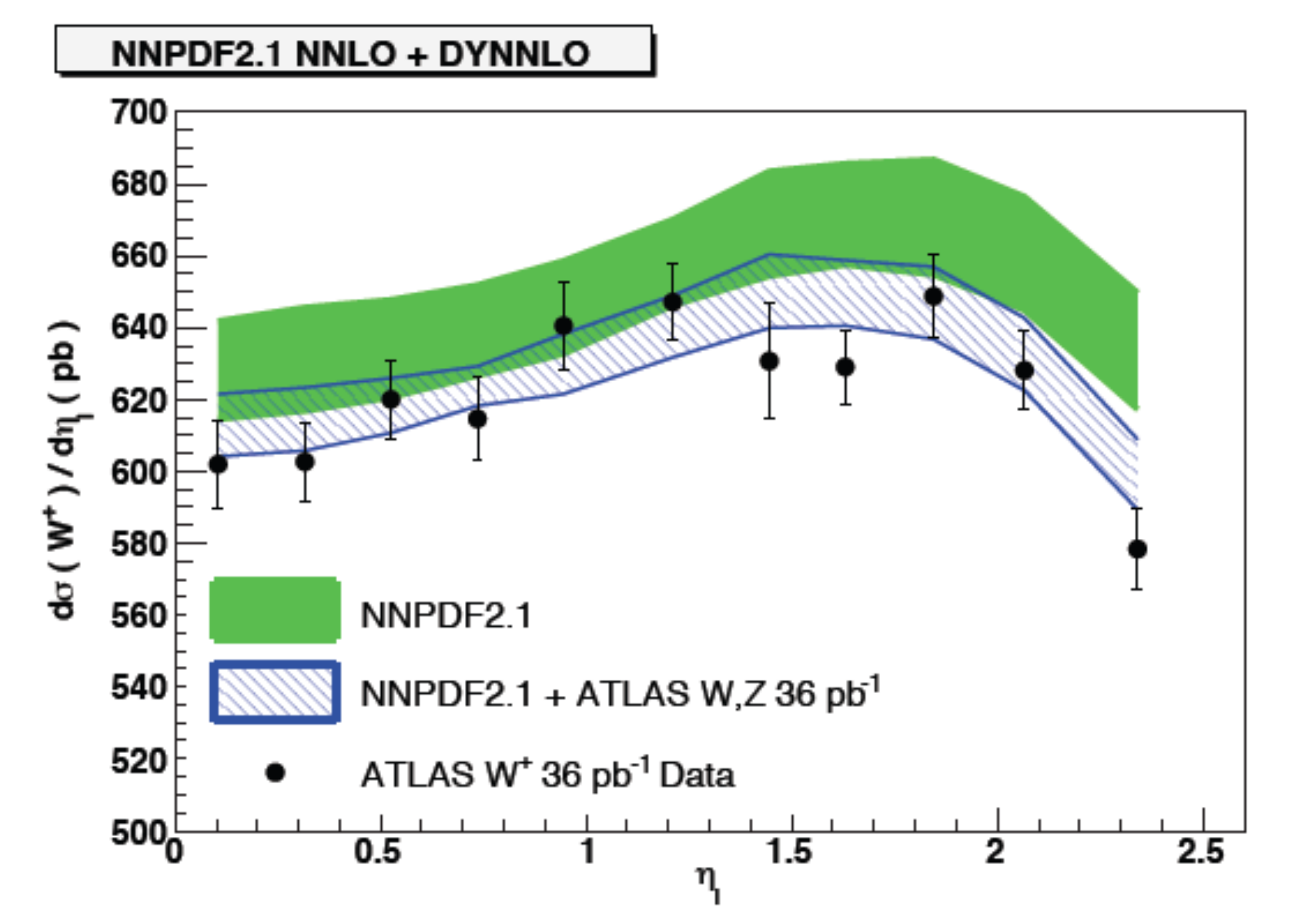}
   \caption{ The ATLAS data on the $W^+$ lepton distribution
from the 2010 data, compared to the NNPDF2.1 NNLO PDFs
before and after including the data into the fit.
}
   \label{fig:atlaswp}
\end{figure}

Each of the LHC processes listed above provides a handle on different parton
combinations and on different $x$-regions. For example,
the light quarks and antiquarks, including the strange PDFs, can be
determined from electroweak boson production. ATLAS has presented a measurement
of the 2010 data for the $W^+$, $W^-$ and $Z$ lepton distributions
with full covariance matrix~\cite{Aad:2011dm}, that can be used to impose tight constraints
on the medium and small-$x$ quarks an antiquarks.
For this dataset the luminosity uncertainties are the dominant experimental
error.
 In Fig.~\ref{fig:atlaswp} we show
how the NNPDF2.1 NNLO prediction are modified by the inclusion of these
ATLAS data, where the theoretical NNLO predictions have been
computed with DYNNLO. 
Related constraints are provided by the CMS  and LHCb $W$
asymmetry measurements~\cite{Chatrchyan:2011jz}, as well
by off--shell Drell--Yan collider data.

\begin{figure}[htb]
   \centering
   \includegraphics*[width=75mm]{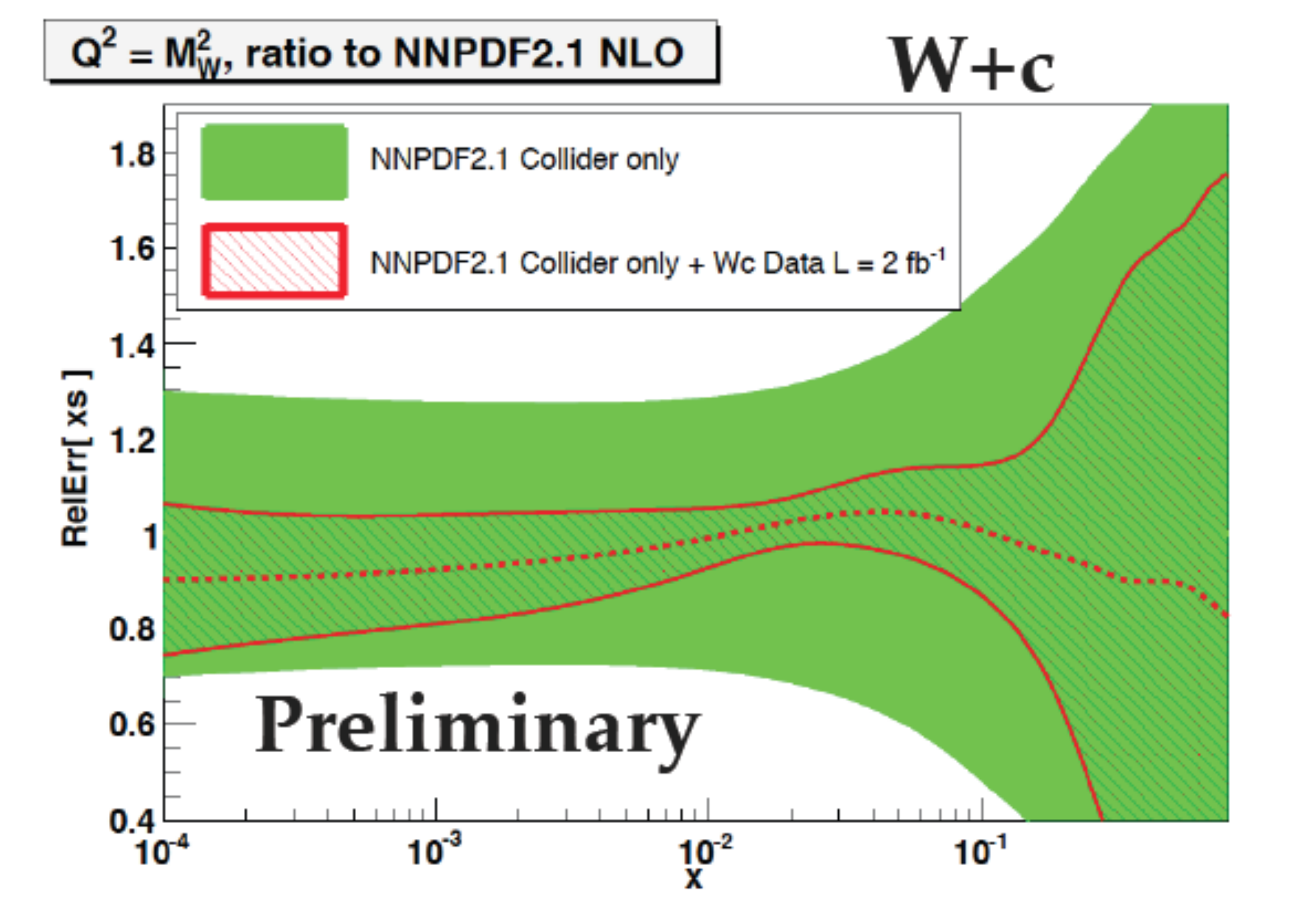}
   \caption{The strangeness distribution
at a typical LHC scale from the NNPDF2.1 NNLO collider
only fit (without LHC data) supplemented by $W+c$ 
pseudodata.
}
   \label{fig:wc}
\end{figure}

Another interesting process is the associated production of
a charm quark together with a $W$ boson:
this process provides a unique direct handle
on the strange quark PDF~\cite{Baur:1993zd}, which is the worse known of all light
quark PDFs. 
In Fig.~\ref{fig:wc} we show preliminary results of a study of
the impact of $W+c$ LHC pseudo-data into the NNPDF2.1 NNLO
collider only PDFs: is clear that the $W+c$ process will allow
a direct determination of the strangeness content of the proton
without the need to resort to low energy neutrino DIS data,
which are less reliable because of large theoretical corrections,
ambiguities of the heavy quark treatment and nuclear corrections.
The inclusive $W$ and $Z$ data should also provide constraints
on the strange sea. A recent ATLAS analysis~\cite{Collaboration:2012sb} 
suggest that their inclusive measurements are accurate enough
to pin down strangeness with good precision, although the analysis
is based on a restrictive parametrization of the input strange PDF.

Other measurements from the LHC are  sensitive to the gluon PDF.
Inclusive jets are known to 
provide an important handle to constrain the large-$x$ gluon.
For example, the D0 and CDF Tevatron Run II measurements are necessary
to stabilize the gluon PDF (as well
as the strong coupling~\cite{Ball:2011us}) in the region relevant for Higgs boson production
via gluon fusion at the LHC~\cite{Thorne:2011kq}. 
ATLAS~\cite{Aad:2011fc}  and 
CMS~\cite{CMS:2011ab}  have presented their results
for inclusive jets and dijets with the 2010 datasets, however only
ATLAS provides the full covariance matrix of the measurement. 
Preliminary results for the CMS inclusive jets and dijets from
the 2011 data,
which span a much wider kinematical range, are also available.

\begin{figure}[htb]
   \centering
   \includegraphics*[width=80mm]{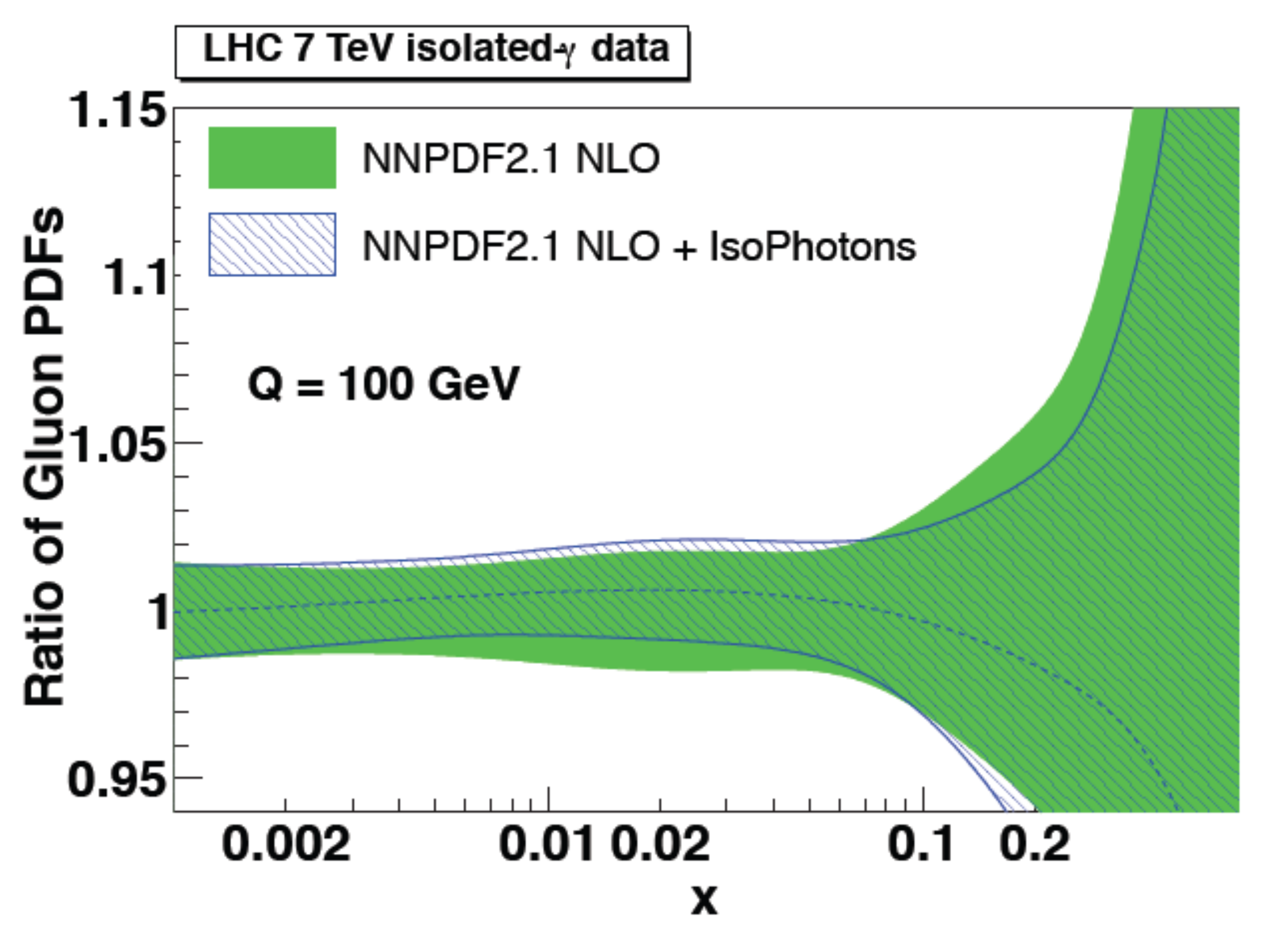}
   \caption{The impact of the recent isolated photon
data from the LHC on the NNPDF2.1 gluon.}
   \label{fig:photons}
\end{figure}

Another process directly sensitive to the gluon PDF at leading
order is direct photon production.
Indeed, prompt photon data was also used until the 90s to
constrain the gluon in PDF fits, but a discrepancy with some fixed
target experiments lead to abandon it. A recent reanalysis shows
a remarkable agreement between NLO QCD and collider isolated
photon data from all scales from 200 GeV up to 7 TeV~\cite{d'Enterria:2012yj},
finding also that the LHC isolated photon data is accurate enough to
provide moderate constraints in the gluon PDF at intermediate values
of $x$, see Fig.~\ref{fig:photons}, and in the associated Higgs
cross sections in gluon fusion.

A careful and consistent treatment of normalization uncertainties
between datasets is crucial in the context of a global PDF 
analysis. Ref.~\cite{Ball:2009qv}  reviews  
the different statistical approaches currently used to
include multiplicative normalization errors into
recent PDF analysis. As an illustration, is clear that it is important
to consistently take common normalizations between
different datasets:
two 2010 ATLAS measurements like inclusive jets
and $W/Z$ production, both available with the
full covariance matrix, have a common single normalization uncertainty
coming from the luminosity calibration, and this is important
for a more accurate PDF analysis. A similar point
was emphasized in Ref.~\cite{Thorne:2011kq}, where it is argued that the
consistent treatment of the normalizations
of  the Tevatron inclusive jet data and Z rapidity
distributions was very important, since the latter fixed the
normalization shift and improved the constraints on the PDFs
coming from the former.

In general, to optimize experimental data combination in a global theoretical 
analysis,
we need to quantify as accurately as possible the possible information:
\begin{itemize}
\item The correlation of the systematic errors, including luminosity, between
two datasets in the same experiment, say $W,Z$ and jet production from ATLAS.
\item The correlation of the systematic errors, including luminosity,  between
two datasets of different experiments, say $W,Z$ from ATLAS and CMS.
\item The correlation of the systematic errors, including luminosity,  between
two datasets of the same experiment from two different runs, say
CMS $W,Z$ between the 2012 and 2011 data.
\end{itemize}
Of course, this is a very optimistic program -- but even if it
can be achieved only in a limited way, it will be very beneficial
for the LHC precision physics studies. Moreover any eventual
combination of LHC data will require to understand this cross-correlations
between different experiments, just as understanding the correlations
between the H1 and ZEUS data lead to the very precise combined
HERA--I dataset~\cite{Aaron:2009aa}.

\section{CROSS SECTION RATIOS AT 8 TEV}

The 8 TeV 2012 LHC run offers a wide new range of physics
possibilities that go beyond the  expanded center
of mass energy. One interesting study  would be to measure ratios
of cross sections between the 2012 and 2011 data.
Such ratios of cross sections are important because they can be measured
very precisely thanks to a strong cancellation of systematic
uncertainties and partly of normalization uncertainties as well.
Thus in principle one can expect to obtain useful information
by selecting suitable observables.

\begin{figure}[htb]
   \centering
   \includegraphics*[width=80mm]{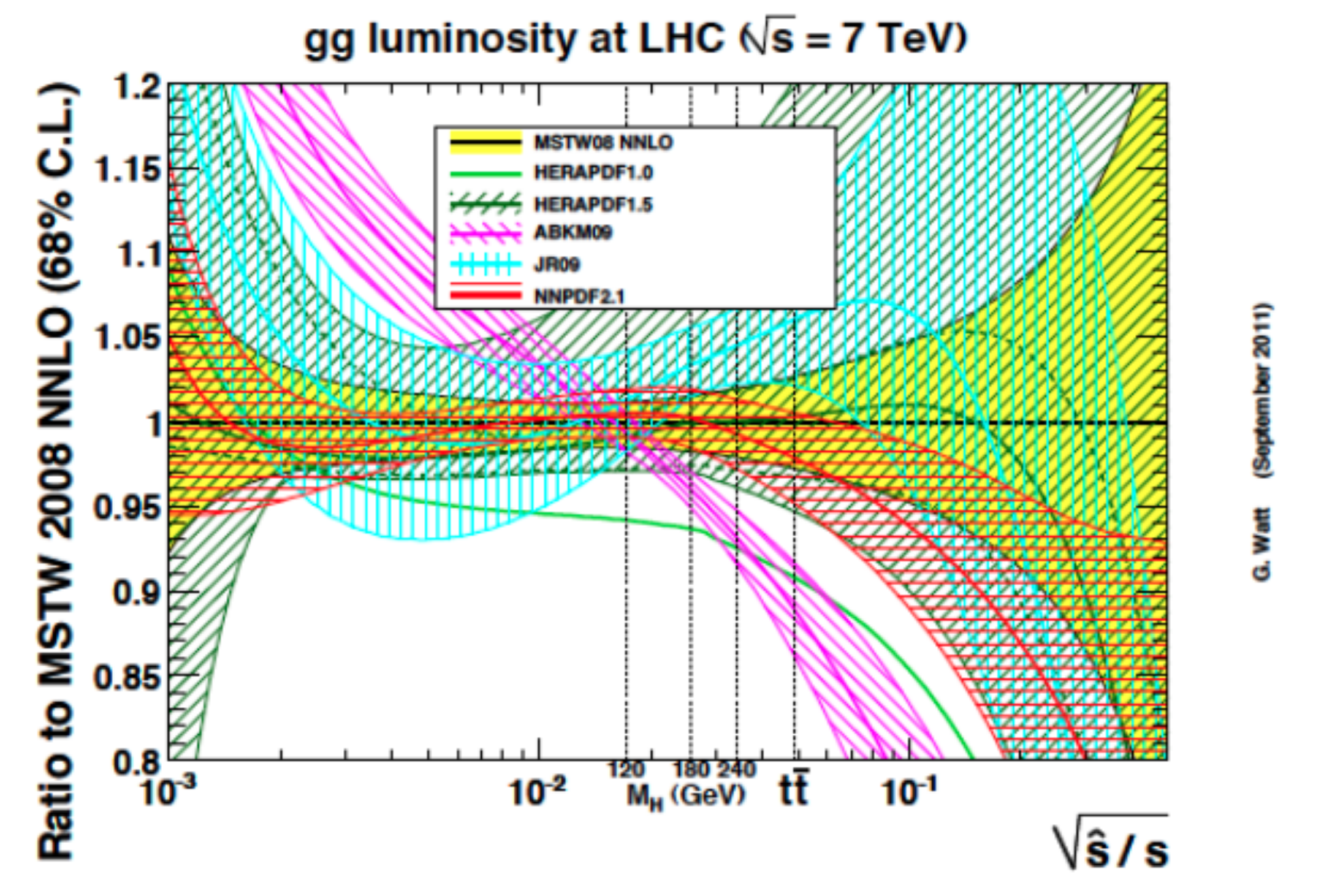}
   \caption{Comparison of the gluon-gluon partonic
luminosity for some recent NNLO PDF sets as a function
of the invariant mass of the produced final state. Note that
PDF uncertainties blow up for very large invariant masses.
Figure taken from~\cite{Watt:2012fj}. }
   \label{fig:gglumi}
\end{figure}

One example is provided by the fact that one can measure ratios
of cross sections between 8 and 7 TeV to constrain the large-$x$ PDFs.
Indeed, large--$x$ PDFs are poorly constrained
by available data: as can be seen Fig.~\ref{fig:gglumi}, the gluon-gluon
parton luminosity when the invariant mass of the produced final state
is large has very substantial uncertainties. This in turn affects
the sensitivity for searches and eventual characterization of new
physics produced close to threshold, like heavy supersymmetric particles
or new resonances.

\begin{figure}[htb]
   \centering
   \includegraphics*[width=80mm]{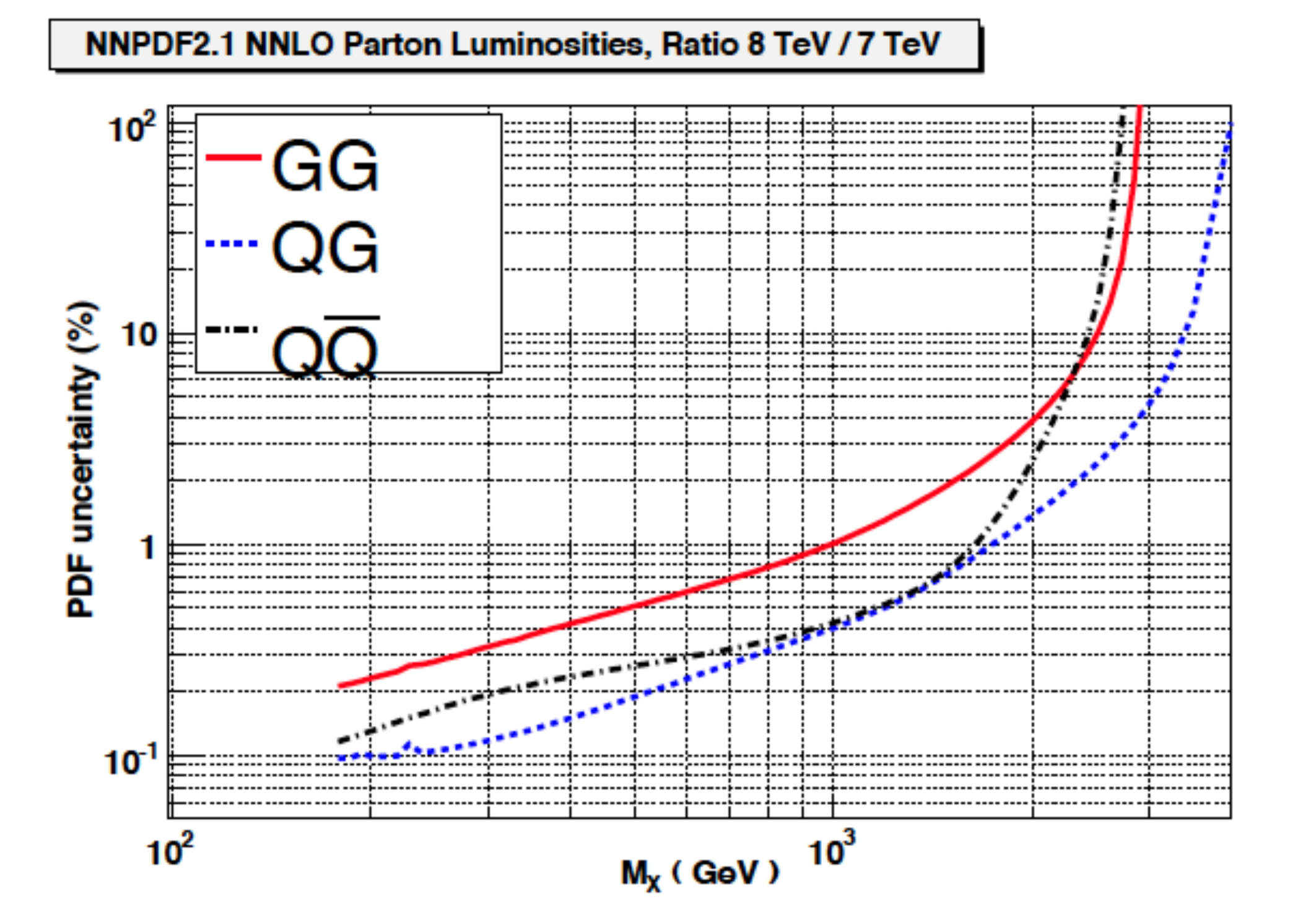}
   \caption{Percentage PDF uncertainty in the ratio of parton luminosities
between 8 TeV and 7 TeV, Eq.~\ref{eq:lumdef}, for the $gg$, $gq$
and $qq$ partonic subchannels, computed with the NNPDF2.1 NNLO set.}
   \label{fig:rojo-lumirat}
\end{figure}

However, it can be seen that large-$x$ PDFs can be studied
by measuring ratios of cross sections at high final state
invariant masses. Indeed, if
one can take the ratio of PDF luminosities between 8 TeV and
7 TeV, for example, the gluon-gluon luminosity 
defined as
\be
R_{\rm gg} \equiv  \frac{\int_{\tau_8}^1
\frac{dx_1}{x_1} g\lp x_1,M_X^2\rp g\lp \tau_8/x_1,M_X^2\rp }{\int_{\tau_7}^1
\frac{dx_1}{x_1} g\lp x_1,M_X^2\rp g\lp \tau_7/x_1,M_X^2\rp}  \ ,
\label{eq:lumdef}
\ee
where $\tau_7=M_X^2/s_7$, $\sqrt{s_7}=7$ TeV and likewise for 8 TeV.
Then one can see that
producing a final state partonic system with invariant mass $M_X$
probes the very large--$x$ PDFs for large $M_X$.
 In Fig.~\ref{fig:rojo-lumirat}
we have computed the 
PDF uncertainty in the ratio of parton luminosities
between 8 TeV and 7 TeV, Eq.~\ref{eq:lumdef} obtained
with the NNPDF2.1 NNLO PDFs~\cite{Ball:2011uy} for different
partonic subchannels.
Is clear that as soon as the mass of the final state object raises
above 1 TeV, the PDF uncertainties on the ratio of luminosities
grows up very fast, up to larger than 100 \%. Thus any measurement
at this scales would provide a major constraint on 
large--$x$
PDFs.

\begin{figure}[htb]
   \centering
   \includegraphics*[width=80mm]{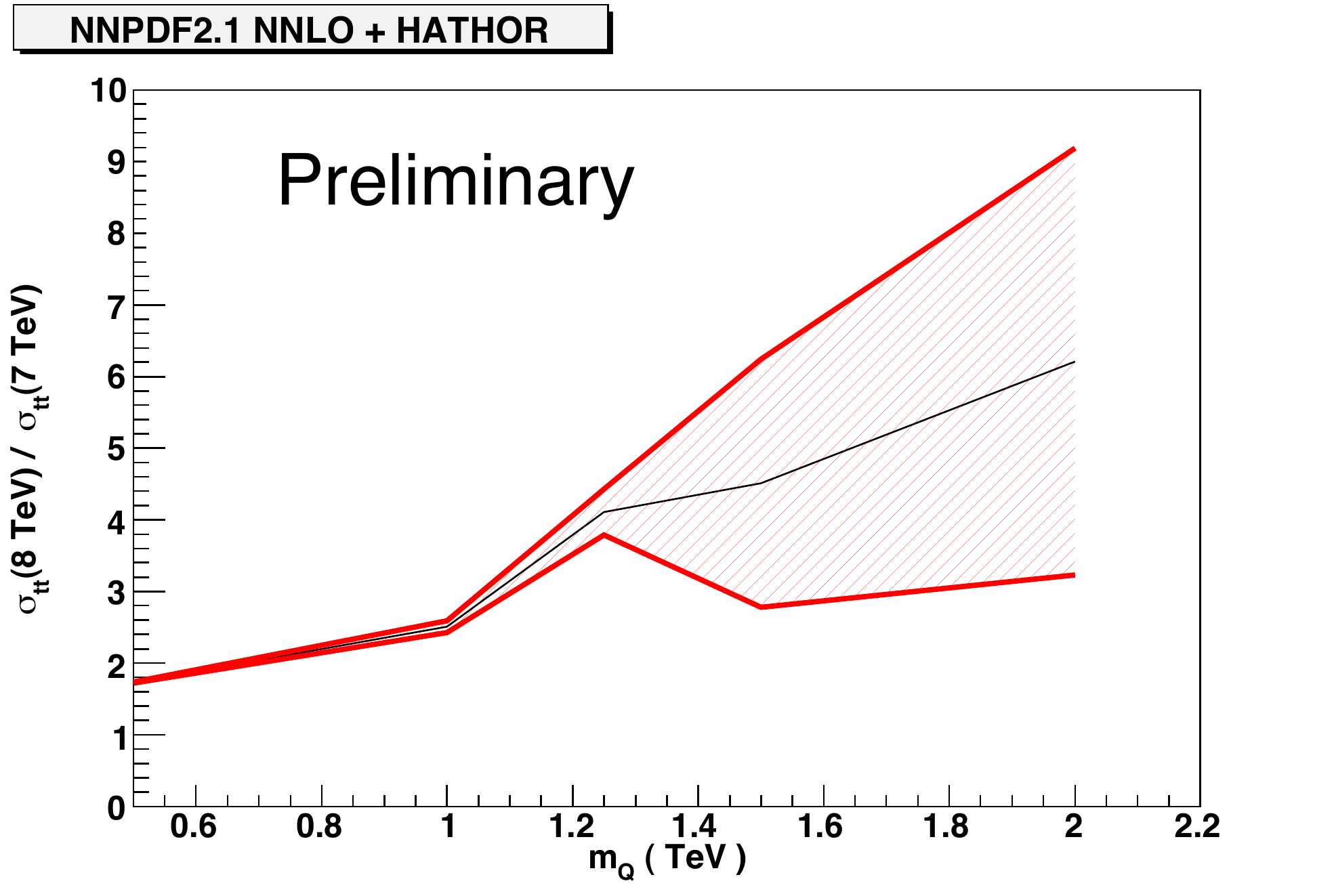}
   \caption{The ratio of production cross sections between
8 and 7 TeV
of a new heavy quark boson with mass $m_Q$ at the TeV scale.
The band represents the PDF uncertainties only. The cross
section has been computed with the HATHOR code and
the NNPDF2.1 NNLO PDFs.}
   \label{fig:hqtop}
\end{figure}

As an illustration of the physics potential of such cross section ratios
measurements,
we show in Fig.~\ref{fig:hqtop} the ratio of production cross sections between
8 and 7 TeV
of a new heavy quark with mass $m_Q$ at the TeV scale.
The band represents the PDF uncertainties only. The cross
section has been computed with the HATHOR code~\cite{Aliev:2010zk}  and
the NNPDF2.1 NNLO PDFs. While PDF uncertainties in the cross
section ratio are small below $m_Q=1$ TeV, above they increase
significantly until they blow up: this is so because of
the cancellation of parton luminosities in Eq.~\ref{eq:lumdef}
fails when the approach the kinematical production threshold.
Therefore, a measurement of a generic cross section ratio with similar
gluon initiated kinematics (like dijet production) would
provide stringent constraints on large-$x$ PDFs even if the
experimental uncertainties are only moderate.

This is just a particular example of how
to use cross section ratios to perform interesting
physics studies, and that can be easily generalized
to many other processes like electroweak production
in association with jets, photon or dijet
production. It is anyway clear that ratios
of cross sections and distributions between 8 and 7 TeV data
have a very interesting physics potential for PDFs, that should be
explored systematically. Many other topics
not necessarily related to PDFs can also be investigated, using
the reduced experimental uncertainties of ratios of cross
sections between different collider energies.

Of course, the effectiveness of cross section ratios data heavily
relies on the possibility to cancel systematic and luminosity
uncertainties in these ratios. The extent to which is possible
still needs to be investigated.

\section{SUMMARY AND OUTLOOK}

A precision measurement of the LHC luminosity is very
important for its physics program. 
 In this contribution we have reviewed the impact
of accurate LHC luminosity determination for 
the data/theory comparison of important standard candles and
then we have discussed the impact of available and future LHC
data in global analysis of parton distributions, with
emphasis on the treatment of luminosity uncertainties.
Finally we have present some possible new physics
opportunities by measuring ratios of cross sections between
th 8 TeV and 7 TeV run, with the particular example
of the determination of large-$x$ parton distributions.

It is clear that understanding to which extent  normalization 
 uncertainties
(as well as in general systematic errors)
between datasets, experiments and runs are correlated
is important to optimize the potential of the LHC data.
In other contributions to these proceedings these
various issues are discussed in more detail. One of the outcomes
of this workshop was that quantifying the correlations of the
luminosity uncertainty between the 2010 and 2011 data was difficult
and that probably they were mostly uncorrelated, however, even
if the correlation is small it will anyway be useful
to quantify it.

To conclude, let us mention that
it was suggested during the workshop to determine the relative
luminosity between ATLAS and CMS by measuring the same process,
say the $Z$ cross section, within the same fiducial volume. 
This is a process for which the dominant error is by far the
normalization uncertainty.
Therefore, it would allow a determination with high accuracy of
the relative luminosity difference between the two experiments,
something which would be an important input to subsequent theoretical
analysis that combine the information from the ATLAS and CMS data.
These kind of studies are very relevant for
the LHC precision program, and we expect that this particular
one and related analysis are carried out in the near future.

\end{document}